\documentclass[12pt]{article}
\renewcommand{\baselinestretch}{1.2}
\usepackage{indentfirst}
\usepackage{amsmath}
\usepackage{amssymb}
\usepackage{amsfonts}
\usepackage{amscd}
\usepackage{amsbsy}
\usepackage{amsthm}
\usepackage{slashed}
\usepackage{bm}
\usepackage{latexsym}
\usepackage{graphicx,color} 
\usepackage[dvipsnames]{xcolor}
\usepackage[colorlinks]{hyperref}
\hypersetup{linkcolor=blue,citecolor=teal,urlcolor=blue}
\def\nn{\nonumber}       
\def\beq{\begin{eqnarray}}
\def\eeq{\end{eqnarray}}
\def\ln{\,\mbox{ln}\,}

\def\det{\,\mbox{det}\,}
\def\tr{\,\mbox{tr}\,}

\def\Tr{\,\mbox{Tr}\,}

\def\sla{\!\!\!\slash}
\def\al{\alpha}
\def\be{\beta}

\def\ga{\gamma}
\def\de{\delta}

\def\ep{\epsilon}
\def\ze{\zeta}

\def\ka{\kappa}
\def\la{\lambda}
\def\na{\nabla}
\def\pa{\partial}

\def\rh{\rho}
\def\si{\sigma}
\def\om{\omega}

\def\ta{\tau}
\def\th{\theta}

\def\Ga{\Gamma}

\def\Si{\Sigma}

\usepackage[symbol]{footmisc} 
\usepackage{float} 
\usepackage{cite}  
\usepackage{titlesec}

\usepackage{ulem} 
\usepackage{orcidlink} 

\titleformat*{\section}{\large\bfseries}
\titleformat*{\subsection}{\normalsize\bfseries}
\begin{document}

\begin{center}
\renewcommand*{\thefootnote}{\fnsymbol{footnote}} 
{\Large \bf
     Extending the Euler-Heisenberg action to include  \\\vspace{0.cm}
 effects of local Lorentz-symmetry violating  \\ \vspace{0.3cm} backgrounds
}
\vskip 6mm

{\bf Wagno Cesar e Silva}\,\orcidlink{0000-0001-7832-3502}
\hspace{-.mm}\footnote{E-mail address: \ wagnorion@gmail.com},
\,
{\bf João Paulo S. Melo}\,\orcidlink{0000-0001-5213-5183}
\hspace{-.mm}\footnote{E-mail address: \ jpsm@cbpf.br}
\,
and
\,
{\bf José A. Helayël-Neto}\,\orcidlink{0000-0001-8310-518X}
\hspace{-.mm}\footnote{E-mail address: \ helayel@cbpf.br}
\vskip 6mm

{
Centro Brasileiro de Pesquisas Físicas,
\\
Rua Dr. Xavier Sigaud 150, Urca, 22290-180, Rio de Janeiro, RJ, Brazil.
}

\end{center}
\vskip 2mm
\vskip 2mm


\begin{abstract}
	
\noindent

This work sets out to compute the corrections to the Euler-Heisenberg 
effective action that arise from spacetime-dependent background 
anisotropies that violate Lorentz symmetry. To accomplish our task, 
we evaluate the functional determinant of the modified Dirac operator 
using the spectral regularization method. Within the framework of the 
Standard Model Extension (SME), these Lorentz-violating parameters 
correspond to the coefficients $m_5(x)$, $a_\mu(x)$, and $b_\mu(x)$. 
The corrected effective action is attained up to the second-order in the 
background parameters. Our results indicate a violation of Furry’s theorem 
at second-order for specific C-odd combinations of these parameters, 
in agreement with previous analyses based on explicit Feynman diagram 
calculations in scenarios with Lorentz-symmetry violation. In the kinetic 
(bilinear) piece of the effective action, non-dynamical axion-like terms 
emerge,  resembling structures commonly encountered in condensed-matter 
systems, such as Weyl semimetals. We show how this procedure modifies 
the Maxwell equations, from which we present both the corresponding 
(non-)conservation of the energy-momentum tensor and the wave equation. 
We also notice that the vacuum behaves as an inhomogeneous medium, and 
compute the local dispersion relation by working in the eikonal approximation. 
As a result of the spacetime-dependence of the background, there appear 
imaginary contributions in the dispersion relation and, consequently, the 
wave amplitudes may be amplified or attenuated, which expresses the 
energy-momentum exchange between the waves and the background.
 
\vskip 3mm


\end{abstract}

\setcounter{footnote}{0} 
\renewcommand*{\thefootnote}{\arabic{footnote}} 

\section{Introduction}
\label{sec1}

Lorentz symmetry is one of the fundamental pillars of modern 
physics, underpinning both special relativity and the framework 
of quantum field theory, which enables the formulation of the 
Standard Model of particle physics. This symmetry guarantees 
the covariance of physical laws under Lorentz transformations, 
ensuring that relativistic field equations preserve their form in all 
inertial frames. Nevertheless, several theoretical developments have 
suggested that this symmetry might not be exact at very high-energy 
scales \cite{Jackiw90, Glashow99, Amelino-Camelia02}.
Possible deviations from Lorentz invariance may emerge, e.g., in the 
context of string-inspired models 
\cite{Kostelecky89, Kostelecky97, Witten99}, quantum gravity 
approaches \cite{Melo24, Pullin99, Nanopoulos99, Urrutia02, Alfaro05}, 
or in anisotropic spacetime constructions \cite{Horava09}.
In such scenarios, Lorentz-symmetry violation gives rise to a broad 
range of nontrivial and far-reaching phenomenological repercussions
\cite{Lane99,  Mewes02, Vargas15, Gomes16, Melo25}.

A systematic framework to investigate the effects of spacetime-symmetry 
breaking is provided by the Standard-Model Extension (SME), 
which describes and classifies all possible Lorentz- and CPT-violating 
terms\footnote{CPT symmetry refers to the combined operation of 
charge conjugation (C), parity transformation (P), and time reversal (T). 
A detailed introduction can be found in many quantum field theory  
textbooks, e.g., in \cite{PeskSchr}.} 
\cite{Kostelecky97, Kostelecky98a, Kostelecky02b, Kostelecky04b}.  
In the Abelian gauge sector, for example, the SME encompasses 
modifications of Maxwell electrodynamics, including the CPT-odd 
Carroll-Field-Jackiw  (CFJ) term 
\cite{Jackiw90, Kostelecky99b, Perez-Victoria01} 
and CPT-even extensions involving a rank-four background tensor with 
the symmetries of the Riemann tensor and a vanishing double trace
\cite{Kostelecky02c}. These modifications may endow the quantum 
vacuum with nontrivial electromagnetic properties, such as birefringence 
or dichroism \cite{Kostelecky98a, Kostelecky02c}.

Within this effective field theory setting, it is also common to consider
extensions of electrodynamics beyond the linear Maxwellian regime. 
In particular, nonlinear electromagnetic theories have been studied in the 
presence of Lorentz-symmetry violation, either as effective descriptions 
arising from spontaneous Lorentz symmetry breaking in Abelian 
gauge-invariant frameworks \cite{Alfaro10}, or as low-energy manifestations 
of quantum gravity-inspired scenarios, such as the semiclassical limit of 
Loop Quantum Gravity \cite{Pullin99, Nanopoulos99, Urrutia02}. 
Related nonlinear structures also appear in brane-world constructions, 
where Dirac-Born-Infeld-type actions may emerge dynamically as a 
consequence of spontaneous Lorentz breaking, even though the primary 
focus lies in the gravitational and brane sectors \cite{Gliozzi11}.
Alternatively, Lorentz symmetry may be explicitly broken in nonlinear 
electrodynamics. An example is provided by Very Special Relativity, where 
a Born-Infeld action can be consistently embedded into a gauge-invariant 
theory with a reduced symmetry group, leading to modified electromagnetic 
interactions, such as a $\tfrac{1}{r^3}$ correction to the Coulomb potential 
\cite{bufalo15}. More recently, phenomenological aspects of nonlinear 
electromagnetic theories coupled to axion fields and Lorentz-violating 
operators have also been investigated in the presence of a CFJ-type 
anisotropic background \cite{Helayel24a}.

Radiative corrections to the Euler-Heisenberg effective action in the 
presence of Lorentz-symmetry violation have been analyzed at one 
loop, typically within setups where the Lorentz-violating backgrounds 
are taken to be constant (see, e.g., \cite{Mariz14}, and references therein). 
In this context, the effects of fermionic operators involving tensorial 
and axial structures were also calculated using the proper-time method 
\cite{Borges16,Ferrari21}. More recently, these investigations were 
extended to scalar QED, yielding nontrivial CPT-even and CPT-odd 
Lorentz-violating contributions in the sector of nonlinear photon 
self-interactions \cite{Petrov25}.

At this stage, a natural question concerns whether more general 
background configurations, in particular those with spacetime 
dependence, can induce physically relevant corrections to these effective 
scenarios. As discussed by Kostelecký in Ref.~\cite{Kostelecky04b}, 
the assumption of constant Lorentz-violating coefficients is not a 
fundamental requirement, but rather a technical simplification. While 
constant backgrounds preserve translation invariance and ensure the 
conservation of the canonical energy-momentum tensor, allowing the 
coefficients to depend on spacetime generally leads to explicit 
modifications of the conservation laws \cite{Helayel24b}. Furthermore, 
in Riemann-Cartan geometries, constant coefficients impose integrability 
conditions that are satisfied only in highly restrictive settings, such as 
parallelizable manifolds, which are of limited physical interest. From this 
perspective, spacetime-dependent Lorentz-violating backgrounds 
constitute a more general and physically well-motivated framework, in 
which small deviations from energy-momentum conservation may arise 
and potentially provide observable signatures of Lorentz symmetry 
violation.

In the present work, we shall address this issue by analyzing the 
additional contributions that arise in the Euler-Heisenberg effective 
framework when Lorentz-violating backgrounds are allowed to 
depend on spacetime. Focusing on the fermionic sector of the SME, 
we compute, via the proper-time method, the complete set of 
corrections to photon-photon scattering induced by 
spacetime-dependent couplings of the form $ a_\mu \gamma^\mu $, 
$ i m_5 \gamma_5 $, and $ b_\mu \gamma^\mu \gamma_5 $. 
Our analysis aims to clarify how such generalized backgrounds 
modify the structure of the effective action and to assess the 
physical consequences of these corrections.

The paper is organized as follows. In Sec.~\ref{sec2}, we briefly 
outline the general structure of the Standard-Model Extension and 
discuss the field redefinitions in the matter sector. Section \ref{sec3} 
is devoted to the calculation of the one-loop effective action,  
including terms up to fourth order in the field strength and second 
order in the  LSV  parameters. In Sec.~\ref{sec4}, we analyze the 
physical implications of the modified effective action. In particular, 
we study the anisotropic corrections to the quadratic sector and 
investigate how Maxwell's equations and the conservation of the 
energy-momentum tensor are modified. Using the eikonal 
approximation we also derive the corresponding local dispersion 
relation. Finally, Sec.~\ref{sec5} contains the concluding remarks. 

We adopt natural units such that $c=\hbar=1$ and use the signature
 $(+,-,-,-)$ for the Minkowski metric $\eta_{\mu\nu}$. 
The paper also includes Appendix~\ref{App}, where we discuss the 
conventions used for the Wick rotation and Euclidean space.

\section{The SME and field redefinitions}
\label{sec2}

The renormalizable and gauge-invariant QED sector of the SME is 
described by the following action in flat spacetime \cite{Kostelecky04b}:
\begin{align}\label{actionnn}
S_{\textrm{\tiny QED-SME}} = 
& 
\int d^4 x \bigg[  \bar{\psi}
\left( i  \Gamma^\mu D_{\mu} -M \right)\psi 
-\dfrac{1}{4 } F^{\mu\nu}F_{\mu\nu}  
-   \dfrac{1}{4 } {( k_{F})}^{\mu\nu\lambda\rho}F
_{\mu\nu}F_{\lambda\rho}  
\nn
\\
&
+  \dfrac{1}{2 }\varepsilon_{\mu\nu\kappa\lambda}
{(k_{AF})}^{\mu}A^{\nu} F^{\kappa\lambda}   \bigg] , 
\end{align} 
where $F_{\mu\nu}=\pa_\mu A_\nu - \pa_\nu A_\mu$ is the 
standard electromagnetic field strength tensor, and 
$D_{\mu} = \pa_\mu +ie A_\mu$ denotes the usual gauge-covariant 
derivative. 
The Lorentz violation\footnote{
	The SME distinguishes observer Lorentz transformations, under which 
	both fields and background coefficients transform covariantly, from 
	particle Lorentz transformations, which act only on dynamical fields 
	and expose the physical breaking of Lorentz symmetry 
	\cite{Kostelecky04b}. Additionally, the CPT theorem establishes that CPT 
	symmetry follows from Lorentz invariance together with the fundamental 
	assumptions of quantum field theory. Consequently, any violation of CPT 
	requires the breakdown of at least one of these conditions, most notably 
	Lorentz invariance. It has been rigorously demonstrated that CPT violation 
	necessarily implies Lorentz violation, whereas Lorentz violation does not, 
	in general, entail CPT violation \cite{Greenberg02, Greenberg06, Lehnert16}.} 
in the gauge sector is characterized by the 
presence of two constant background quantities. The first corresponds 
to a CPT-even contribution, represented by the dimensionless tensor 
${(k_{F})}^{\mu\nu\lambda\rho}$, which exhibits the same symmetry 
properties as the Riemann curvature tensor in the gravitation context. 
The second is a CPT-odd term, described by the vector 
${(k_{AF})}^{\mu}$, which has canonical mass dimension and 
generates the well-known CFJ term. In the matter sector, the 
Lorentz-violating terms are given by
\begin{align}
\Gamma^\mu 
&
=  
\gamma^\mu + c^{\mu\nu}\gamma_\nu 
+d^{\mu\nu}\gamma_\nu\gamma_5 
+e^\mu 
+  i f^\mu\gamma_5  
+ \dfrac{1}{4}g^{\lambda\nu\mu}\Sigma_{\lambda\nu} 
\label{GGterm} 
\end{align}
and
\begin{align}
M 
&
= m 
+ im_5 \gamma_5 
+ a_\mu \gamma^\mu 
+ b_\mu \gamma^\mu\gamma_5  
+ H_{\mu\nu}\Sigma^{\mu\nu}.
\label{MMterm}
\end{align}
The Dirac matrices $\gamma^\mu$ satisfy the Clifford algebra 
$\left\{\gamma^\mu, \gamma^\nu \right\} = 2\eta^{\mu\nu}$, 
with the chirality matrix defined by 
$\gamma_5 = i\gamma^0\gamma^1\gamma^2\gamma^3$, 
and the generators of the proper orthochronous Lorentz group 
$SO(1,3)$ in the spinor representation are written as 
$\Sigma^{\mu\nu} = \tfrac{1}{2}\sigma^{\mu\nu}=
\tfrac{i}{4}[\gamma^\mu,\gamma^\nu]$.
Generally, the structures in Eqs.\,\eqref{GGterm} and 
\eqref{MMterm} can be expressed, respectively, as 
$\Gamma^\mu = \gamma^\mu 
+ \delta \Gamma_{\textrm{\tiny LSV}}^\mu$ and 
$M = m + \delta M_{\textrm{\tiny LSV}}$, where 
$\delta \Gamma_{\textrm{\tiny LSV}}^\mu$ and 
$\delta M_{\textrm{\tiny LSV}}$ denote small and constant Lorentz- 
and CPT-violating corrections. 

Requiring the Lagrangian to be Hermitian implies that all LSV 
coefficients must be real. The parameters $ a_\mu $, $ b_\mu $, 
$ e_\mu $, and $ g_{\mu\nu\lambda} $  are CPT-odd, whereas 
$ f_\mu $ is CPT-even. This latter property follows from the fact 
that $ f_\mu $ can be absorbed into the coefficients $ c_{\mu\nu} $ 
by means of an appropriate spinor redefinition, so that $ f_\mu $ 
contributes to physical observables only through bilinear 
combinations $ f_\mu f_\nu $ \cite{Altschul06}. It is also worth 
emphasizing that, in certain cases, a field redefinition can be 
performed to eliminate some of the background terms appearing in 
Eqs.\,\eqref{GGterm} and \eqref{MMterm}. In other words, some 
of the LSV-terms can be absorbed into a reparametrization of the 
spinor field $\psi$. A typical example is the $ a_\mu $ term, 
which can be removed by the transformation 
$\psi(x) = e^{-i a \cdot x} \chi(x)$. Similarly, the $ b_\mu $ 
term could also be eliminated through a suitable redefinition, 
but only in the absence of the fermionic mass term, $ m $. 
Another LSV-term that can be straightforwardly removed is $ m_5 $, 
through a chiral rotation of the form 
$\psi = e^{i\alpha \gamma_5} \psi'$, with a global parameter given 
by $\alpha = -\tfrac{1}{2} \tan^{-1} \left( \tfrac{m_5}{m} \right)$, 
and a corresponding redefinition of the mass term: 
$m \to m' = m \cos(2\alpha) - m_5 \sin(2\alpha)$ \cite{Altschul06} 
(see also \cite{Colladay02b}, for a rigorous analysis of field 
redefinitions that can eliminate redundant background choices).

Regarding the Lorentz and CPT invariance of the 
action~\eqref{actionnn}, it is important to recall that these 
symmetries are defined with respect to the free field theory, 
$\mathcal{L}_0 = \bar{\psi}(i \gamma^\mu \partial_\mu - m)\psi$. 
From this perspective, the distinction between observer and 
particle transformations becomes essential. In a certain sense, 
Lorentz symmetry remains preserved: the theory behaves normally 
under rotations and boosts as long as one considers passive 
Lorentz transformations, i.e., those associated with changes of 
coordinates in the laboratory frame. 
Within the framework of the SME, LSV emerges exclusively under 
active transformations, when particle transformations modify the 
orientation of the fields relative to fixed background tensor 
structures that characterize the vacuum. In this formulation, 
the LSV terms, such as $ (k_{F})^{\mu\nu\lambda\rho} $, 
$ (k_{AF})^{\mu} $, $ \delta\Gamma_{\textrm{\tiny LSV}}^\mu $, 
and $ \delta M_{\textrm{\tiny LSV}} $, transform as tensors, 
vectors, or scalars under observer transformations, but remain 
invariant (i.e., behave effectively as scalars) under particle 
transformations. This feature enables these terms to be 
consistently treated as constant background fields 
\cite{Kostelecky97}. Furthermore, it is worth noting that the 
pseudoscalar mass parameter $ m_5 $ does not introduce an 
apparent violation of Lorentz symmetry, since the combined 
presence of $ m $ and $ m_5 $ still yields a Lorentz-invariant 
mass term at the level of the dispersion relation 
\cite{Helayel24b, Kostelecky04b}.

In what follows, we shall restrict our analysis only to the 
Lorentz symmetry-violating corrections in the fermionic mass 
sector, \( \delta M_{\textrm{\tiny LSV}} \), setting all 
\( H_{\mu\nu}  \), \( (k_{F})^{\mu\nu\lambda\rho}  \), 
\( (k_{AF})^{\mu}  \), and 
\( \delta\Gamma_{\textrm{\tiny LSV}}^\mu \) 
equal to zero. Furthermore, we assume that the background 
responsible for Lorentz violation is spacetime-dependent. 
Under these assumptions, the modified covariant QED action in 
Minkowski spacetime, incorporating Lorentz-symmetry breaking 
contributions, reads
\begin{align}\label{Action_1}
S_{1} = \int d^4 x \bigg[ 
- \dfrac{1}{4}F_{\mu\nu} F^{\mu\nu}
+ \bar\psi (i \gamma^\mu D_\mu 
-m 
-im_5\gamma_5
-a_\mu \gamma^\mu 
-b_\mu \gamma^\mu \gamma_5
)\psi \bigg],
\end{align}
where now $a_\mu=a_\mu(x)$, $b_\mu=b_\mu(x)$ and 
$m_5=m_5(x)$. In this case, the spinor field redefinitions 
discussed above are no longer sufficient to eliminate $a_\mu(x)$ 
and $m_5(x)$. The $a_\mu$-term cannot be removed, since the 
derivative operator acts on both $x_\mu$ and $a_\mu$ in the 
exponential factor $e^{-ia\cdot x}$, thereby generating additional 
derivative couplings. On the other hand, the elimination of the 
$m_5$-term through a chiral rotation leads to a 
spacetime-dependent mass, which complicates the interpretation 
of the resulting parameter as the physical fermion mass. 
Therefore, in the setting \eqref{Action_1}, all parameters $m_5$, 
$a_\mu$, and $b_\mu$ are expected to provide nontrivial 
contributions to Lorentz symmetry breaking and to play a 
significant role in the analysis presented throughout this work.

\section{One-loop effective action}
\label{sec3}

In this section, based on the spectral regularization method 
\cite{McGar97}, we employ a procedure similar to that adopted 
in Refs. \cite{Borges16,Ferrari21} to evaluate the functional 
determinant of the modified Dirac operator. Since our interest is 
to integrate out only the matter fields (fermions, in this case), 
while keeping the gauge field as a classical background, the 
starting point is the fermionic functional trace associated with 
the action \eqref{Action_1},
\begin{align} \label{quadraticO}
-i\Tr\ln\big(i \slashed{D}-\mathcal{M}\big) = -
\frac{i}{2}\Tr\Big[\ln\big(i \slashed{D}-\mathcal{M}\big)
+\ln\big(-i \slashed{D} -\mathcal{M}^\star\big)\Big] 
= -\frac{i}{2}\Tr\ln \mathcal{O},
\end{align}
where $\Tr$ denotes the functional trace, including integration 
over spacetime coordinates and trace over spinor indices. For 
convenience, we introduce the compact notation
\begin{align} \label{MandMstar}
\mathcal{M} \equiv m 
+im_5\gamma_5
+a_\mu \gamma^\mu 
+b_\mu \gamma^\mu \gamma_5,
\qquad
\mathcal{M}^\star \equiv m 
+im_5\gamma_5
-a_\mu \gamma^\mu 
-b_\mu \gamma^\mu \gamma_5,
\end{align}
and adopt the Feynman slash notation 
$\slashed{D} \equiv \ga^\mu D_\mu$.
The squared operator $\mathcal{O}$ can be written in the form
\begin{align}\label{dec_O}
	\mathcal{O} 
	&
	= \mathcal{K}\Big[1+\mathcal{K}^{-1}\mathcal{X}\Big],
\end{align}
with $\mathcal{K} \equiv D^2+m^2+e\,\Si^{\mu\nu}F_{\mu\nu}$ 
representing the minimal second-order differential operator, 
and the Lorentz-violating part given by
\begin{align}\label{LVpart}
	\mathcal{X} 
	\equiv
	& 
	\;-a^2+b^2-m_5^2
	+i\big(\pa\cdot a+2\,a\cdot D\big)
	+i\ga_5\big(\pa\cdot b
	+2m_5 m\big)
	+2im_5\slashed{b}
	-\gamma_5\big(
	\slashed{\pa} m_5
	\nn 
	\\ 
	&
    \;
	+2m_5 \slashed{D}
	+2im_5 \slashed{a} \big)
	+4\Si^{\mu\nu}\big(
	2ia_{[\mu} b_{\nu]} \gamma_5
	-2\gamma_5 b_{[\mu} D_{\nu]} 
	-\pa_{[\nu} a_{\mu]}
	-\ga_5\,\pa_{[\nu} b_{\mu]}\big).
\end{align}
Since the Lorentz-violating parameters are expected to be small 
(see, e.g., Ref. \cite{Kostelecky97} for theoretical discussions), the 
decomposition \eqref{dec_O} proves particularly useful, as it allows 
for a systematic expansion of the effective action in which 
$\mathcal{X}$ can be treated as a small perturbation around the 
minimal operator $\mathcal{K}$,
\begin{align}\label{func_trace}
	\Ga 
	=
	-\frac{i}{2}\Tr\ln \mathcal{O}
	& 
	= -\frac{i}{2}\Tr\ln \mathcal{K}
	-\frac{i}{2}\Tr \big(\mathcal{K}^{-1}\mathcal{X}\big)
	+\frac{i}{4}\Tr \big(\mathcal{K}^{-1}\,\mathcal{X}
	\,\mathcal{K}^{-1}\,\mathcal{X}\big)
	+ (\ldots),
\end{align}
where $(\ldots)$ denotes higher-order terms in $\mathcal{X}$. In the 
present work, we restrict our analysis to contributions up to second 
order in the Lorentz-violating background parameters. In this regard, 
it is important to note that the third term in Eq.\,\eqref{func_trace} 
contains cubic and quartic powers of the Lorentz-violating structures 
and, hence, these contributions will be neglected in what follows. 

In order to perform the analytic continuation from Minkowski to Euclidean 
space, we implement a Wick rotation of the time coordinate, 
$x^0 \to -i\hat x_4$, in the squared operator $\mathcal{O}$. Consequently, 
the gauge-covariant derivative in Euclidean space takes the form\footnote{
	Additional details on the conventions employed for the Wick rotation of 
	the remaining operator structures can be found in Appendix \ref{App}.}
\begin{align} \label{ECcovder}
	\hat D_\mu=\hat \pa_\mu-i e \hat  A_\mu,
\end{align}
and the effective action becomes
\begin{align}
     \hat \Ga = -i \Ga.
\end{align}

Then, by collecting only the first- and second-order terms in the 
Lorentz-violating parameters, and expressing the inverse of the minimal 
operator via the Schwinger proper-time representation, 
\begin{align} \label{propertime}
	\frac{1}{\mathcal{\hat{K}}}=\int_0^\infty ds\,e^{-s\mathcal{\hat{K}}},
\end{align}
we arrive at
\begin{align} \label{EucldianActions}
	\hat \Ga 
	=
	\hat \Ga_0+ \hat \Ga_{\textrm{LSV}},
\end{align}
where $\hat \Ga_0$ denotes the Lorentz-invariant part of the effective 
action, corresponding to the standard QED contribution \cite{Shwinger51}, 
and $\hat \Ga_{\textrm{LSV}}$ encodes the contributions induced by the 
Lorentz-violating background.  In what follows, for notational simplicity, 
we omit the hats on quantities defined in Euclidean space. 
In this way, at first-order in the Lorentz-violating parameters we have
\begin{align}\label{LSV1}
\Ga_{\textrm{LSV}}\big|^{(1)}
=&
-\frac{i}{2\mu^2}\Tr \int_0^\infty ds \, \Big\{
e^{\frac{s}{\mu^2}(D^2-m^2-e\,\Si^{\la\si}F_{\la\si})}
\Big[
\pa\cdot a+2\,a\cdot D
+\ga_5\big(\pa\cdot b+2m_5 m\big)
\nn 
\\ 
&
+2i\Si^{\mu\nu}\big(
\pa_{\nu} a_{\mu}
+\ga_5\,\pa_{\nu} b_{\mu}
+2\gamma_5 b_{\mu} D_{\nu} \big)
\Big]
\Big\},
\end{align}
while the second-order terms are
\begin{align}\label{LSV2}
\Ga_{\textrm{LSV}}\big|^{(2)}
=&\;
-\frac{1}{2\mu^2}\Tr \int_0^\infty ds \, \Big\{
e^{\frac{s}{\mu^2}({{D}}^2-m^2-e\,\Si^{\la\si}{F}_{\la\si})}
\Big[
a^2-b^2+m_5^2
-4i\Si^{\mu\nu}\ga_5 a_{\mu} b_{\nu}
\Big]
\Big\}
\nn 
\\ 
&\;
-\frac{1}{4\mu^4}\Tr \int_0^\infty ds\int_0^\infty d\ta \,\Big\{
e^{\frac{s}{\mu^2}({{D}}^2-m^2-e\,\Si^{\la\si}{F}_{\la\si})}\,
\Big[
\pa\cdot a+2\,a\cdot D
+\ga_5\big(\pa\cdot b
\nn 
\\ 
&\; +2m_5 m\big)
+2i\Si^{\mu\nu}\big(
\pa_{\nu} a_{\mu}
+\ga_5\,\pa_{\nu} b_{\mu}
+2\gamma_5 b_{\mu} D_{\nu} \big)
+i\ga_5\big({\pa}\sla m_5
+2m_5\slashed{D}\big) 
\Big]
e^{\frac{\ta}{\mu^2}{{D}}^2}
\nn 
\\ 
&\;
\times e^{-\frac{\ta}{\mu^2}{(m^2+e\,\Si^{\ga\rh}{F}_{\ga\rh})}}
\Big[
\pa\cdot a+2\,a\cdot D
+\ga_5\big(\pa\cdot b+2m_5 m\big)
+2i\Si^{\al\be}\big(
\pa_{\be} a_{\al}
\nn 
\\ 
&\;
+\ga_5\,\pa_{\be} b_{\al}
+2\gamma_5 b_{\al} D_{\be} \big) +i\ga_5\big({\pa}\sla m_5
+2m_5\slashed{D}\big)
\Big]
\Big\}.
\end{align}
Here, $\mu$ denotes a mass scale introduced to render the 
proper-time parameter dimensionless. Notice that in 
Eqs.~\eqref{LSV1} and \eqref{LSV2} we have already discarded 
terms containing an odd number of $\gamma$-matrices, whose 
traces vanish identically.

In what follows, we evaluate these contributions separately, since 
the expressions are lengthy and technically involved, particularly 
in the second-order case. 

\subsection{First-order LSV corrections}
\label{sec3.1}

Considering the functional trace, defined as 
$\Tr = \tr \textstyle \displaystyle{\int} d^4x\,\de^4(x-x')\big|_{x=x'}$\,, 
and making use of the trace cyclic property to change 
$e^{-\frac{\ta}{\mu^2}{{D}}^2}$ position, the expression \eqref{LSV1} 
reads
\begin{align}
	\Ga_{\textrm{LSV}}\big|^{(1)}
	=&\; 
	-\frac{i}{2\mu^2} \int d^4x \int_0^\infty ds \, \Big\{
	e^{-\frac{sm^2}{\mu^2}}
	\tr e^{-\frac{s}{\mu^2}(e\,\Si^{\la\si}{F}_{\la\si})}
	\Big[
	\pa\cdot a+2\,a\cdot D
	+\ga_5\big(\pa\cdot b+2m_5 m\big)
	\nn 
	\\ 
	&\;
	+2i\Si^{\mu\nu}\big(
	\pa_{\nu} a_{\mu}
	+\ga_5\,\pa_{\nu} b_{\mu}
	+2\gamma_5 b_{\mu} D_{\nu} \big)
	\Big]
	e^{\frac{s}{\mu^2}{{D}}^2}
	\de^4(x-x')\big|_{x=x'}
	\Big\}.
\end{align}
The kernel can be expressed as \cite{McGar97}
\begin{align}
	K(s)
	& 
	=
	e^{\frac{s}{\mu^2}D^2}\de^4(x-x')\big|_{x=x'} 
	=
	\frac{\mu^4}{16\pi^2 s^2}
	\det^{1/2}\bigg[\frac{iesF}{\mu^2\sinh\big(\frac{ies}{\mu^2}F\big)}\bigg].
\end{align}
It is worth noting that the terms containing a single gauge-covariant 
derivative acting on the kernel vanish, since \cite{McGar97}
\begin{align}
	D^\mu e^{\frac{s}{\mu^2}D^2}\de^4(x-x')\big|_{x=x'} 
	& 
	=
	0.
\end{align}
Then,
\begin{align}\label{intLSV1}
	\Ga_{\textrm{LSV}}\big|^{(1)}
	=&\;	
   - \frac{i\mu^2}{32\pi^2} \int d^4x \int_0^\infty ds \, \bigg\{\frac{1}{s^2}
	\,e^{-\frac{sm^2}{\mu^2}}
	\tr e^{-\frac{s}{\mu^2}(e\,\Si^{\la\si}{F}_{\la\si})}
	\Big[
	\pa\cdot a
	-\ga_5\big(\pa\cdot b-2m_5 m\big)
	\nn 
	\\ 
	&\;
	+2i\Si^{\mu\nu}\big(\pa_{\nu} a_{\mu}
	-\ga_5\,\pa_{\nu} b_{\mu}\big)
	\Big]
	\det^{1/2}\bigg[\frac{iesF}{\mu^2\sinh\big(\frac{ies}{\mu^2}F\big)}\bigg]
	\bigg\}.
\end{align}
For weak background electromagnetic fields, we can proceed with 
the following expansion:
\begin{align}\label{expKernel}
	\det^{1/2}\bigg[\frac{iesF}{\mu^2\sinh\big(\frac{ies}{\mu^2}F\big)}\bigg]
	& 
	=
	1+\frac{e^2s^2}{12\mu^4}\Tr(F^2)
	+\frac{e^4s^4}{72\mu^8}\bigg[\frac14 F^4
	+\frac15 \Tr(F^4)\bigg]+O(F^6).
\end{align}
In this way, the traces of the exponentials are explicitly given by
\begin{align}\label{serie1}
	\tr \Big[e^{-\frac{s}{\mu^2}(e\,\Si^{\la\si}F_{\la\si})}\Big]
	=
	& 
	\,\tr 
	\bigg[\de^\mu_\mu
	-\frac{es}{\mu^2}\,\Si^{\la\si}F_{\la\si}
	+\frac{e^2s^2}{2\mu^4}\,\big(\Si^{\la\si}F_{\la\si}\big)^2
	-\frac{e^3s^3}{6\mu^6}\,\big(\Si^{\la\si}F_{\la\si}\big)^3
	\nn
	\\
	&
	+\frac{e^4s^4}{24\mu^8}\,\big(\Si^{\la\si}F_{\la\si}\big)^4
	+\ldots
	\bigg]
	\nn
	\\
	=
	&
	\;4
	+\frac{e^2 s^2}{\mu^4} F^2
	+\frac{e^4 s^4}{2\mu^8} \bigg(\frac14F^4
	-\frac13F^{\mu\nu}F_{\nu\al}F^{\al\be}F_{\be\mu}\bigg)
	+O(F^6);
\end{align}
\begin{align}\label{serie2}
	\tr \Big[e^{-\frac{s}{\mu^2}(e\,\Si^{\la\si}F_{\la\si})}\ga_5\Big]
	=
	& 
	\,-\frac{e^2 s^2}{\mu^4}(\tilde{F}^{\mu\nu}F_{\mu\nu})
	-\frac{e^4 s^4}{12\mu^8}F^2(\tilde{F}^{\mu\nu}F_{\mu\nu})
	+O(F^6);
\end{align}
\begin{align}\label{serie3}
	\tr \Big[e^{-\frac{s}{\mu^2}(e\,\Si^{\la\si}F_{\la\si})}\Si^{\mu\nu}\Big]
	=
	& 
	\,-\frac{2e s}{\mu^2}F^{\mu\nu}
	-\frac{e^3 s^3}{\mu^6}\bigg[\frac12F^2F^{\mu\nu}
	+\frac{2}{3}F^{\mu\la}F_{\la\si}F^{\si\nu}\bigg]
	+O(F^5);
\end{align}
and
\begin{align}\label{serie4}
	\tr \Big[e^{-\frac{s}{\mu^2}(e\,\Si^{\la\si}F_{\la\si})}\Si^{\mu\nu}\ga_5\Big]
	=
	& 
	\,\frac{2e s}{\mu^2}\tilde F^{\mu\nu}
	+\frac{e^3 s^3}{12\mu^6}
	\Big[F^2\tilde F^{\mu\nu}
	+\tilde F_{\la\si}F^{\la\si} F^{\mu\nu}
	+2\tilde F_{\la\si}F^{\mu\la}F^{\si\nu}
	\nn
	\\
	&
	-F^{\mu\la} F_{\la\si} \tilde F^{\si\nu} 
	-\tilde F^{\mu\la}F_{\la\si}F^{\si\nu}\Big]
	+O(F^5),
\end{align}
with $\tilde F_{\mu\nu}=\tfrac{i}{2}  \varepsilon_{\mu\nu\al\be} F^{\al\be}$ 
denoting the dual field strength tensor in the Euclidean space.

By inserting the expressions derived above into \eqref{intLSV1},  evaluating 
the proper-time integrals, and returning to the Minkowski spacetime, we 
arrive at the following result
\begin{align}\label{order1}
\Ga_{\textrm{LSV}}\big|^{(1)} =
 -\frac{e^2}{16\pi^2m} \int d^4x \, m_5 
\tilde{F}^{\mu\nu}F_{\mu\nu}\,,
\end{align}
where we neglect contributions that, in the covariantly constant background 
field approximation, reduce to total derivative terms after integration by parts.

One notes that the axial mass parameter $m_5(x)$ gives rise to a nontrivial 
contribution to the effective action at first-order in LSV, with the same local 
structure as an axion-photon coupling. In contrast, for constant \(m_5\), this 
contribution reduces to a surface term, up to integration by parts, and hence 
does not lead to physical modifications of the electromagnetic dynamics. 
Nevertheless, even in the spacetime-dependent case, \(m_5(x)\) should be 
regarded as an external pseudoscalar background rather than a propagating 
axion-like field, since no kinetic or potential terms for $m_5$ appear at this 
order.
 
 An interesting point concerns the possible relevance of terms analogous to 
 the action \eqref{order1} in condensed-matter systems. Nonpropagating 
 excitations, in some cases spacetime-dependent and effectively described 
 by axion-like parameters, have been reported in several low-energy scenarios 
 (see, e.g., \cite{Nenno20} for the review). Particularly noteworthy are systems 
 in which such excitations arise in materials exhibiting emergent Lorentz 
 symmetry (and its violation \cite{kostelecky22}), as discussed in Refs. 
 \cite{Gooth19,Sekine21,Grigoreva25}.
In this context, Euler-Heisenberg-type nonlinear excitations have also been 
reported in these materials \cite{Keser22}. Therefore, approaches such as the 
one developed in the present work may provide a suitable framework for 
describing these systems, in which multiple effects can coexist.

 Since the action \eqref{order1} is the only term contributing at first-order in 
 LSV, we show in what follows that additional axion-like terms induced by 
 Lorentz breaking emerge at second-order in LSV corrections.

\subsection{Second-order LSV  corrections}
\label{sec3.2}

We now turn to the evaluation of the second-order contributions to Lorentz 
violation, encoded in $\Ga_{\textrm{LSV}}\big|^{(2)}$. The first line of 
\eqref{LSV2} contains only a proper-time integral and can thus be computed 
in a straightforward way, by applying the same procedures used in the 
derivation of the first-order LSV corrections. In particular, one makes use of 
the cyclic property of the trace and the expansions given in 
Eqs. \eqref{expKernel} -- \eqref{serie2}, leading to
\begin{align}\label{int2}
	\Ga_{\textrm{LSV}}\big|^{(2)}
	& 
	=
	-\frac{1}{2\mu^2} \int d^4x \int_0^\infty ds \, \Big\{
	e^{-\frac{sm^2}{\mu^2}}
	\tr e^{-\frac{s}{\mu^2}(e\,\Si^{\la\si}{F}_{\la\si})}
	\Big[
	a^2-b^2+m_5^2
	-4i\Si^{\mu\nu}\ga_5 a_{\mu} b_{\nu}
	\Big]
	K(s)
	\Big\}
	\nn 
	\\ 
	&
	-\frac{1}{4\mu^4} \int d^4x \int_0^\infty ds\int_0^\infty d\ta \,\Big\{
	e^{-\frac{(s+\ta)m^2}{\mu^2}}
	\tr e^{-\frac{s}{\mu^2}(e\,\Si^{\la\si}{F}_{\la\si})}\,
	\Big[
	\pa\cdot a+2\,a\cdot D
	+\ga_5\big(\pa\cdot b
	\nn 
	\\ 
	&
	+2m_5 m\big)
	+2i\Si^{\mu\nu}\big(
	\pa_{\nu} a_{\mu}
	+\ga_5\,\pa_{\nu} b_{\mu}
	+2\gamma_5 b_{\mu} D_{\nu} \big)
	+i\ga_5\big({\pa}\sla m_5
	+2m_5\slashed{D}\big)
	\Big]
	e^{-\frac{\ta}{\mu^2}(e\,\Si^{\ga\rh}{F}_{\ga\rh})}
	\nn 
	\\ 
	&
	\times
	e^{\frac{\ta}{\mu^2}{{D}}^2}
	\Big[
	\pa\cdot a+2\,a\cdot D
	+\ga_5\big(\pa\cdot b+2m_5 m\big)
	+2i\Si^{\al\be}\big(
	\pa_{\be} a_{\al}
	+\ga_5\,\pa_{\be} b_{\al}
	+2\gamma_5 b_{\al} D_{\be} \big)
	\nn 
	\\ 
	&
	+i\ga_5\big({\pa}\sla m_5
	+2m_5\slashed{D}\big)
	\Big]
	e^{-\frac{\ta}{\mu^2}{{D}}^2}
	e^{\frac{(s+\ta)}{\mu^2}{{D}}^2}\de^4(x-x')
	\Big\}.
\end{align}

On the other hand, the part of Eq.\,\eqref{int2} containing two proper-time 
integrals involves more complicated operator structures. Their treatment 
requires the use of nontrivial commutation relations between derivative 
operators and background fields. To evaluate such commutators, we 
employ the identities derived from the Baker-Campbell-Hausdorff series:
\begin{align}
	e^{\frac{\ta}{\mu^2}{D}^2}\big({\pa}_\mu b^\mu
	+2 b^\mu {D}_\mu\big)e^{-\frac{\ta}{\mu^2}{D}^2} 
	=
	&
	\;{\pa}_\mu b^\mu+2 b^\mu {D}_\mu +
	\frac{\ta}{\mu^2} [{D}^2,{\pa}_\mu b^\mu+2 b^\mu {D}_\mu]
	\nn
	\\
	& 
	+ \frac{\ta^2}{2\mu^4}[{D}^2,[{D}^2,{\pa}_\mu b^\mu+2 b^\mu {D}_\mu]]
	+\ldots
	\nn
	\\
	=
	& 
	\;\big(e^{-\frac{2ie\ta}{\mu^2}{F}}\big)_\mu^{\;\;\,\nu}\,\big({\pa}_\nu b^\mu
	+2 b^\mu {D}_\nu\big)
	+({\pa}_\nu b^\mu)\Pi_\mu^{\;\;\nu}\,,
\end{align}
with
\begin{align}
	\big(e^{-\frac{2ie\ta}{\mu^2}{F}}\big)_\mu^{\;\;\,\nu}\,({\pa}_\nu b^\mu
	+2 b^\mu {D}_\nu)
	=
	&
	\;
	\bigg[\de^\nu_\mu
	+\frac{2 i \ta e}{\mu^2} F_{\mu}^{\;\;\nu}
	-\frac{2\ta^2e^2}{\mu^4}F_{\mu\la}F^{\la\nu}
	-\frac{4i \ta^3e^3}{3\mu^6} F_{\mu\la}F^{\la\ga}F_\ga^{\;\;\nu}
	\nn
	\\
	& 
	+\frac{2 \ta^4e^4}{3\mu^8} F_{\mu\la}F^{\la\ga}F_{\ga\ze}F^{\ze\nu}
	+O(F^5)
	\bigg]({\pa}_\nu b^\mu+2 b^\mu {D}_\nu),
\end{align}
and
\begin{align}
	\Pi_\mu^{\;\;\nu}
	=
	&
	\;\frac{2\ta}{\mu^2}\bigg\{
	2 D_\mu D^\nu
	+i eF_\mu^{\;\;\nu}
	+\frac{ie\ta}{\mu^2}\Big[4F_{\mu\la} D^\la D^\nu+2F^{\nu\la} D_\la D_\mu
	+3ieF_{\mu\la}F^{\la\nu}\Big]
	\nn
	\\
	& 
	-\frac{2e^2 \ta^2}{3\mu^4}\Big[
	6 F_{\mu\la}F^{\la\ga}D_\ga D^\nu
	+ 2 F^{\nu\la}F_{\la\ga} D^\ga D_\mu
	+ 6 F_{\mu\la}F^{\nu\ga}D_\ga D^\la
	-5ieF_{\mu\la}F^{\la\ga}F_{\ga}^{\;\;\nu}\Big]
	\nn
	\\
	& 
	-\frac{i e^3\ta^3}{3\mu^6}\Big[
	8F_{\mu\la}F^{\la\ga}F_{\ga\ze}D^\ze D^\nu
	+2F^{\nu\la}F_{\la\ga}F^{\ga\ze} D_\ze D_\mu
	-5ieF_{\mu\la}F^{\la\ga}F_{\ga\ze}F^{\ze\nu} 
	\nn
	\\
	& 
	+4 \big(3F_{\mu}^{\;\;\la}F^{\nu}_{\;\;\ga}
	+2F_{\mu\ga}F^{\nu\la}\big) F_{\la\ze} D^\ze D^\ga
	\Big]
	+\frac{4e^4\ta^4}{15\mu^8}\Big[
	5F_{\mu\la}F^{\la\ga}F_{\ga\ze}F^{\ze\xi}D_\xi D^\nu
	\nn
	\\
	& 
	+F^{\nu\la}F_{\la\ga}F^{\ga\ze}F_{\ze\xi} D^\xi D_\mu
	+5 \big(2F_{\mu}^{\;\;\la}F^{\nu}_{\;\;\ga}F_{\la}^{\;\;\ze}
	+F_{\mu\ga}F^{\nu\la}F_{\la}^{\;\;\ze}
	+2F_{\mu}^{\;\;\ze}F^{\nu\la}F_{\la\ga}\big) F_{\ze\xi} D^\xi D^\ga
	\Big]
	\nn
	\\
	& 
	+O(F^5)
	\bigg\}.
\end{align}
Another useful relation is
\begin{align}
	e^{\frac{\ta}{\mu^2}{D}^2}b^\mu e^{-\frac{\ta}{\mu^2}{D}^2} 
	=
	&
	\;
	b^\mu
	+\frac{2\tau}{\mu^2}(\pa_\ga b^\mu)\bigg[
	\de_\rh^\ga
	+\frac{i e\tau}{\mu^2} F^{\ga}_{\;\;\rh}
	-\frac{2e^2\ta^2}{3\mu^4} F^{\ga\si}F_{\si\rh}
	-\frac{ie^3\ta^3}{3\mu^6} F^{\ga\si}F_{\si\ze}F^{\ze}_{\;\;\rh}
	\nn
	\\
	&
	+\frac{2e^4\ta^4}{15\mu^8} F^{\ga\si}F_{\si\ze}F^{\ze\xi}F_{\xi\rh}
	+O(F^5)\bigg]D^\rh
	\nn
	\\
	=
	&
	\;b^\mu+(\pa_\ga b^\mu)\bar\Pi^\ga_{\;\;\rh} D^\rh.
\end{align}
It is worth emphasizing that these relations are valid for a background with 
constant field strength, i.e., $[{D}_\mu,{D}_\nu]=-ie{{F}}_{\mu\nu}=$ constant. 
 Furthermore, for simplicity, we restrict our analysis to Lorentz-violating backgrounds $m_5(x)$, $a_\mu(x)$, and $b_\mu(x)$ that depend linearly on the spacetime coordinates, as adopted for the gauge field $A_\mu(x)$ in the constant-field-strength approximation. This choice corresponds to the simplest nontrivial class of spacetime-dependent Lorentz-violating backgrounds.\footnote{
 Consequently, all terms involving second and higher-order derivatives of the Lorentz-violating backgrounds vanish identically. In this framework, one may also introduce field-strength-like quantities associated with the Lorentz-violating parameters, which are likewise constant. 
}
Using these results, we can proceed to compute the relevant contributions 
to the effective action, which are given by
\begin{align}
	\Ga_{\textrm{LSV}}\big|^{(2)}
	&
	=
	-\frac{ \mu^2}{32\pi^2 } \int \! d^4x \! \int_0^\infty \!\! ds  
	\frac{1}{s^2}e^{-\frac{sm^2}{\mu^2}}
	\bigg\{
	\bigg[4
	+\frac{e^2 s^2}{\mu^4} F^2
	+\frac{e^4 s^4}{2\mu^8} \bigg(\frac14F^4
	-\frac13F^{\mu\nu}F_{\nu\al}F^{\al\be}F_{\be\mu}\bigg)
	\bigg]
	\nn
	\\
	&
	\times \big(a^2+b^2-m_5^2\big)
	-i\frac{8e s}{\mu^2}\tilde F^{\mu\nu}a_\mu b_\nu
	-i\frac{e^3 s^3}{3\mu^6}
	\Big[F^2\tilde F^{\mu\nu}
	+\tilde F_{\la\si}F^{\la\si} F^{\mu\nu}
	+2\tilde F_{\la\si}F^{\mu\la}F^{\si\nu}
	\nn
	\\
	&
	-F^{\mu\la} F_{\la\si} \tilde F^{\si\nu}
	-\tilde F^{\mu\la}F_{\la\si}F^{\si\nu}\Big]a_\mu b_\nu 
	\bigg\}
	-\frac{1}{4\mu^4} \int d^4x \int_0^\infty ds\int_0^\infty d\ta \,\bigg\{
	e^{-\frac{(s+\ta)m^2}{\mu^2}}
	\nn 
	\\ 
	&
	\times 
	\tr e^{-\frac{s}{\mu^2}(e\,\Si^{\la\si}{F}_{\la\si})}\,
	\Big[
	2i\Si^{\mu\nu}\big(\pa_{\nu} a_{\mu}
	-\ga_5\,\pa_{\nu} b_{\mu}\big)
	-\ga_5\big(\pa\cdot b
	+2\,b\cdot D
	-2m_5 m\big)
	+\pa\cdot a
	\nn 
	\\ 
	&
	+2\,a\cdot D
	-\ga_5\big({\pa}\sla m_5
	+2m_5\slashed{D}
	-2m\slashed{b}\big)
	\Big]
	\,e^{-\frac{\ta}{\mu^2}(e\,\Si^{\ga\rh}{F}_{\ga\rh})}\,
	\bigg[
	2i\Si^{\mu\nu}\big(\pa_{\nu} a_{\mu}
	-\ga_5\,\pa_{\nu} b_{\mu}\big)
	\nn 
	\\ 
	&
	+\big(e^{\frac{2ie\ta}{\mu^2}{F}}\big)_\mu^{\;\;\,\nu}\,
	\big({\pa}_\nu a^\mu
	+2 a^\mu {D}_\nu
	-\ga_5\pa_\nu b^\mu-2\ga_5b^\mu {D}_\nu
	-\ga_5\ga^\mu\pa_\nu m_5-2\ga_5\ga^\mu m_5D_\nu\big)
	\nn 
	\\ 
	&
	+\big(\pa_\nu a^\mu
	-\ga_5\pa_\nu b^\mu
	-\ga_5\ga^\mu\pa_\nu m_5\big)\Pi_\mu^{\;\;\nu}
	+2\ga_5m\big(m_5+\slashed{b}\big)
	+2\ga_5m\big(\pa_\ga m_5
	+\pa_\ga \slashed{b}\big)\bar\Pi^\ga_{\;\;\rh}D^\rh
	\bigg]
	\nn 
	\\ 
	&
	\times\,e^{\frac{(s+\ta)}{\mu^2}D^2}\de^4(x-x')
	\bigg\},
\end{align}
where the result of the action of two gauge-covariant derivatives on 
the kernel can be expressed as \cite{McGar97}
\begin{align}
D^\mu D^\nu e^{\frac{(s+\ta)}{\mu^2}D^2}\de^4(x-x')\big|_{x=x'} 
=
& 
\,\bigg[\frac{-ieF}{e^{2ie(s+\ta)F/\mu^2}-1}\bigg]^{\nu\mu} K(s+\ta)
\nn
\\
=
&
\,\bigg[-\frac{\mu^2}{2(s+\ta)}\de^{\mu\nu}
-\frac{i e}{2}F^{\mu\nu}
+\frac{e^2(s+\ta)}{6\mu^2}F^{\mu\ka}F_{\ka}^{\;\;\nu}
\nn
\\
&
\,+\,\frac{e^4(s+\ta)^3}{90\mu^6}F^{\mu\ka}F_{\ka\th}F^{\th\eta}F_{\eta}^{\;\;\nu}
\bigg] K(s+\ta).
\end{align}
The terms involving two proper-time integrals require extensive Clifford 
algebra manipulations, in particular due to the presence of products of 
several Lorentz generators in combination with $\ga_5$. For this reason, 
the computation of these structures was carried out separately, order by 
order in the field strength, and independently cross-checked using the 
xAct package in \textit{Mathematica} \cite{Wolfram}. After evaluating the 
corresponding traces, performing the integrations over the proper 
times $s$ and $\ta$, and returning to Minkowski spacetime,  one arrives at
\begin{align}
	\Ga_{\textrm{LSV}}\vert^{(2)} = \sum_{n=1}^4  \Ga_{F^n},
\end{align}
where $n$ labels the power of the field strength appearing in each
term of the effective action. For $n = 1$, we then obtain corrections 
linear in $F$ and second-order in the LSV background, given by
\begin{align} \label{FTviolation1}
	\Ga_{F^1} = \dfrac{ e}{16\pi^2 m }\int d^4 x \,
	\left\{ \left[\dfrac{1}{\epsilon} 
	- \ga_E + \ln \left( \dfrac{\mu^2}{m^2} \right) \right]2 a_\mu b_\nu \tilde F^{\mu\nu}    
	+  2m_5(\partial_\mu a_\nu )\tilde{F}^{\mu\nu} 
	\right\} , 
\end{align}
where $\ga_E $ is the Euler-Mascheroni constant. The term 
$a_\mu b_\nu F^{\mu\nu}$ in the action \eqref{FTviolation1} originates 
from a purely divergent contribution. This corresponds to shifting the vacuum, 
since this infinity contribution appears in the tadpole of the photon field. 
In contrast, the term proportional to $m_5 (\partial_\mu a_\nu)\tilde{F}^{\mu\nu}$ 
is finite and suggests that LSV background can effectively act as a 
source term in Maxwell equations.

For $n = 2$, the contribution takes the following form
\begin{align} \label{mazwellmod1}
	\Ga_{F^2} =&\; \dfrac{e^2}{16 \pi^2m^2} \int d^4 x \, 
	\bigg[ 
	\dfrac{1}{3} \big ( m^2_5+ b^2 \big )F^2 
	+ \dfrac{1}{6} (a \cdot b) F_{\mu\nu} \tilde{F}^{\mu\nu}       
	\bigg].
\end{align}
 In this quadratic sector of the effective action, the second-order corrections 
 in LSV contribute to the Maxwell equations in a nontrivial way. It is important 
 to note that, as in the case of the contribution to the photon kinetic sector 
 given by Eq.\,\eqref{order1}, the new terms appearing in Eq.\,\eqref{mazwellmod1} 
 also contain axion-like contributions. However, it should be emphasized that 
 these are not dynamical axions, but rather axion-type parameters that may 
 be related to possible phenomenological applications.

Now, for $n = 3$, we obtain that
\begin{align} \label{FTviolation3}
	\Ga_{F^3} =&\; \dfrac{ e^3}{16\pi^2 m^4} \int d^4 x \,
	\Bigg\{
	\dfrac{37}{96}  a^\mu b^\nu   F_{\mu\nu} (F \cdot\tilde F)    
	-\dfrac{7}{5} \dfrac{m_5}{m} \big( \pa^\mu a^\nu \big) F_{\mu\nu}
	\bigg[F^2  -\dfrac{1}{2}(F\cdot\tilde F) \bigg]   \bigg\} .
\end{align}
Finally, but not least, we obtain the contribution for $n = 4$, given by
\begin{align} \label{ac4ligh-light}
	 \Ga_{F^4} 
    &=  -
    \dfrac{ e^4}{16\pi^2 m^6} \int d^4 x
    \,\bigg\{ 
    \dfrac{m_5^2}{4}\bigg[ 
    \dfrac{1}{9} F^4 
    +\dfrac{7}{45}    \big( F \cdot \tilde F \big)\bigg]   
    +\dfrac{523}{5472} \big(a \cdot b\big)  F^2 \big (F \cdot \tilde F \big) 
    \nn
    \\
    &
    +  \dfrac{b^2}{45} \bigg[  F^4 + 2   \big( F \cdot \tilde F \big)^2\bigg]
    +\dfrac{2}{45} b^\mu b_\nu  F_{\mu\ka} F^{\ka\nu } F^2 
    - \dfrac{817}{720}a^\mu b_\nu F_{\mu\ka} F^{\ka\nu} \big(F \cdot \tilde F\big)  
    \nn
    \\
    &
    - \dfrac{m_5}{m} \bigg[  \dfrac{31}{60} 
     \big( \pa^\mu a_\nu \big) F_{\mu\ka} F^{\ka\nu}    \big( F \cdot \tilde F \big)  
    +\dfrac{2}{25} \big( \pa \cdot b \big)   \big( F \cdot \tilde F \big)^2  
    -12 \big( \pa^\mu b_\nu \big)  F_{\mu\ka} F^{\ka\nu} F^2  \bigg]
    \bigg\} .
\end{align}

The results \eqref{FTviolation1} and \eqref{FTviolation3} point to a violation of 
Furry’s theorem. This theorem states that any Feynman diagram containing a 
closed fermion loop with an odd number of external photon lines gives a 
vanishing contribution to physical amplitudes \cite{Furry51}. This occurs because 
the QED interaction is invariant under charge conjugation, while the photon field 
changes sign under this transformation. As a consequence, diagrams related 
by reversing the direction of the fermion loop cancel each other exactly when 
the number of photon vertices is odd. Furry’s theorem plays an important role 
in simplifying perturbative calculations in QED and explains, e.g., why processes 
such as single-photon or three-photon amplitudes generated by a fermion loop 
do not contribute to the effective action. 

The seminal work \cite{Lane02} studies the validity of Furry’s theorem in the 
presence of first-order corrections in LSV and shows that only C-odd fermionic sectors are expected to yield contributions that violate the theorem. In our case, where the LSV backgrounds are spacetime dependent, the terms responsible for the violation of Furry's theorem arise from the mixing of two distinct LSV sectors and appear only at second-order. This
indicates that the results on the violation of Furry’s theorem obtained in 
Ref.~\cite{Lane02} can be extended to second-order contributions in LSV.

Finally, we obtain the second-order LSV contribution to photon-photon scattering, 
given by Eq.\,\eqref{ac4ligh-light}. Actually, this set of quartic terms in the field strengths of the photon can be discussed in the context of the anomalous vertices in the electroweak  theory. Moreover, the $Z^0$ field also couples, through weak coupling constant,   to electron-positron field so that as similar neutral four-vertices can also appear associated $Z^0$ field \cite{Kubota14,Kupco18}. This means that they may be used to constrain the corresponding  parameters through photon-photon scattering data from the ATLAS and CMS  experiments \cite{ATLAS17, ATLAS19, CMS19}.  Also,  we should point out that quartic terms cast in Eq.\,\eqref{ac4ligh-light} yield tree-level contributions to the photon-photon scattering. 

At this point, it is worth commenting on the absence of CFJ-type terms, of the form 
$\sim A_\mu \tilde{F}^{\mu\nu}$. In Ref.~\cite{Jackiw99}, it was shown that the 
calculation of the vacuum polarization with first-order insertions of the term 
$b_\mu \gamma^\mu \gamma_5$ and constant $b_\mu$, leads to a CFJ-type 
correction to the effective action at second-order in the photon field, with  $(k_{AF})^\mu = \tfrac{3}{16\pi^2} b^\mu$. However, it was emphasized that 
this result depends on the regularization scheme employed.  In our calculation, the electromagnetic background is frozen, $F_{\mu\nu}=\text{constant}$, hence avoiding contributions involving
higher-derivative and CFJ-type terms. Once we freeze the field strength, it is equivalent to adopting the
Fock-Schwinger gauge, $A_\mu=-\tfrac12 F_{\mu\nu}x^\nu$, and consequently  it becomes clear
that the effective action is written entirely in terms of $F_{\mu\nu}$. There is no way the gauge potential explicitly appears.  

A additional remark concerns the fact that, if we restrict ourselves to the regime in which the coefficients are constant, in addition to discarding terms involving derivatives acting on the LSV background, we must also neglect the contributions depending on $a_\mu$ and $m_5$ at all orders in the field strength of the corrected effective action. This follows from the fact that one can perform a field redefinition from the outset to eliminate these terms, indicating that they cannot generate physical   effects, as discussed in Sec. \ref{sec2}.

In this regard, by setting the parameters 
$a_\mu$ and $m_5$ to zero and assuming $b_\mu$ to be constant in 
Eq.\,\eqref{mazwellmod1}, one obtains 
\begin{align}\label{bb_term}
	\Gamma_{F^2} = 
    \dfrac{e^2}{48 \pi^2 m^2} \int d^4 x \, b^2 F^2.
\end{align}
In contrast to the result presented in work~\cite{Ferrari21}, when considering 
in Eq.\,\eqref{bb_term} the particular case in which $b_\mu$ is light-like, we 
find that the quadratic photon sector does not receive contributions from 
second-order in LSV. This discrepancy stems from the use of different 
parametrizations in the construction of the quadratic operator given in 
Eq.\,\eqref{quadraticO}. In the present work, we make use of the properties 
of the $\gamma_5$ matrix together with the cyclicity of the trace, whereas 
Ref.~\cite{Ferrari21} adopts a parametrization that effectively changes the 
sign of the mass term $m$ and of $b_\mu \ga^\mu \gamma_5$. 
As a consequence, a relative sign difference in the mass term $m$ inside 
of $\mathcal{M}^*$, Eq.\,\eqref{MandMstar}, appears when comparing the 
two approaches. 

This ambiguity in the parametrization of the quadratic Dirac operator is 
related to the so-called multiplicative anomaly (see, e.g., 
\cite{Dowker,Evans,Sh09} for the details). In a work currently in 
preparation \cite{JPWH26}, we revisit the calculation of the effective action 
with contributions from two constant Lorentz-violating parameters, $b_\mu$ 
and $H_{\mu\nu}$, and analyze this issue in a fully general scenario, with 
arbitrary parametrization. 
In the present work, we choose to use the parametrization implemented 
through the $\gamma_5$ matrix, since it is the only one consistent with 
all results available in the literature obtained via explicit Feynman-diagram 
loop calculations in presence of LSV, such as those reported in 
\cite{Lane02, Gomes10}.

In the next section, we analyze how the quadratic sector, composed of 
the results \eqref{order1} and \eqref{mazwellmod1}, modifies Maxwell's 
equations and the conservation of the energy-momentum tensor. We 
also derive the local dispersion relation associated with this corrections.

\section{Modified Maxwell equations} \label{sec4}
The kinetic sector that contributes to vacuum Maxwell equations is given by
\begin{align}
    \Gamma_{\tiny \textrm{kinetic}} 
   = \Gamma_{\tiny \textrm{Maxwell}} 
   + \Gamma_{\textrm{LSV}}\big|^{(1)}   
   + \Gamma_{F^2} 
   =-\dfrac{1}{4} \int d^4 x \, \Big[ 
   \big(1+c_1\big) F^2 +(c_2 +c_3)  \tilde{F}^{\mu\nu}F_{\mu\nu} 
   \Big],
\end{align}
where we have used  the definition of the  fine-structure constant  
$\al = \tfrac{e^2}{4\pi}$ and the following compact notations:
\begin{align}
    c_1 &= -\dfrac{\al}{3\pi m^2} \big(m_5^2+  b^2 \big) , \\
    c_2 &= -\frac{\al}{6\pi m^2}      (a \cdot b )
      , \\
    c_3 &= \dfrac{\al}{\pi m} m_5 \,.
\end{align}
 From the variational principle applied to the photon field, we obtain the 
 covariant Maxwell equations in the form
\begin{align} \label{CovariantMEq}
 &\big(1+c_1\big) \pa_\mu F^{\mu \nu } 
 +  \big( \pa_\mu c_1\big) F^{\mu\nu} 
 + \big[ \pa_\mu \big( c_2 + c_3\big) \big]  \tilde F^{\mu\nu }= 0 , 
\end{align}
where the Bianchi identity $\partial_\mu \tilde F^{\mu\nu} = 0$ holds. 
Since LSV is expected to be small, Eq.\,\eqref{CovariantMEq} can be 
rewritten as $\pa_\mu F^{\mu \nu }  = \big(1+c_1 \big)^{-1} (...) $, 
allowing for a perturbative expansion of the right-hand side in powers 
of $c_1$. Keeping terms up to second-order in the LSV parameters,
Eq.\,\eqref{CovariantMEq} reduces to
\begin{align} \label{CovariantMEq2}
 \pa_\mu F^{\mu\nu} = &\;  \big(\pa_\mu c_1 \big) F^{\mu\nu}  
 + \big[ \pa_\mu \big( c_2 + c_3\big) \big] 
    \tilde{F}^{\mu\nu}  .
\end{align}
The energy-momentum tensor conservation can be explicitly verified 
by contracting the right-hand side of Eq.\,\eqref{CovariantMEq2} with 
$F_{\nu\rho}$ and subsequently constructing a continuity equation of 
the form $\partial_\mu (\dots)$. This procedure leads to
\begin{align} \label{EMtensor}
    \pa_\mu \Big[(1+c_1) {T^\mu}_\ka\Big] 
    = (1+c_1) J^\mu F_{\mu\ka} 
    + \dfrac{1}{4} (\pa_\ka c_1) F_{\al\be}F^{\al\be} 
    + \dfrac{1}{4}\big[ \pa_\ka \big( c_2 + c_3\big) \big]  F_{\al\be}\tilde F^{\al\be} , 
\end{align}
where ${T^\mu}_\kappa = F^{\mu\nu}F_{\nu\kappa} 
+ \tfrac{1}{4}\delta^\mu_\kappa F_{\alpha\beta}F^{\alpha\beta}$ 
is the standard energy-momentum tensor of Maxwell 
electrodynamics. In Eq.\,\eqref{EMtensor}, the source term $J^\nu$ 
may arise either from the coupling of the photon field to matter or 
from a purely LSV-induced source term originating from the effective 
action ~\eqref{FTviolation1}, when it is divergence-free. We observe 
that allowing the LSV background to depend on the spacetime 
coordinates leads to a violation of the conservation of the 
energy-momentum tensor, which is restored in the limit of a 
constant background.

This non-conservation of the energy-momentum tensor requires a careful assessment of the domain of validity of the effective description. It is well known that spacetime-dependent Lorentz-violating backgrounds generally lead to the non-conservation of the energy-momentum tensor \cite{Kostelecky04b} and, in certain cases, obstruct the consistent definition of a symmetric energy-momentum tensor \cite{Helayel24b,Spallicci18}. In the present case, this feature arises from treating the coefficients $m_5(x)$, $a_\mu(x)$, and $b_\mu(x)$ as nondynamical external backgrounds. Consequently, Lorentz symmetry is broken explicitly, and the apparent non-conservation of the electromagnetic energy-momentum tensor can be interpreted as an exchange of energy and momentum with the background sector.

On the other hand, as discussed in Refs.~\cite{Kostelecky89,Kostelecky97,Kostelecky04b}, energy-momentum conservation is recovered once the complete dynamical system is taken into account. In scenarios with spontaneous Lorentz-symmetry breaking, the Lorentz-violating coefficients arise as vacuum expectation values of underlying dynamical fields, so that the apparent non-conservation in the observable sector reflects an exchange of energy and momentum with the fields responsible for the symmetry breaking, while the total energy-momentum tensor remains conserved.

Furthermore, Refs.~\cite{Kostelecky04b,Bluhm05,Bluhm24} establish a series of no-go results showing that explicit local Lorentz-symmetry breaking is incompatible with the geometric framework of gravitational theories formulated on Riemann and Riemann-Cartan manifolds. In such geometries, diffeomorphism invariance together with the Bianchi identities impose constraints on the gravitational field equations that require the covariant conservation of the energy-momentum tensor, a condition naturally satisfied in the case of spontaneous symmetry breaking. Consequently, spacetime-dependent nondynamical backgrounds may still be consistently employed in flat spacetime as an effective description, provided that the theory is not regarded as a closed system. Therefore, the no-go results constrain the gravitational extension of such constructions rather than their use as effective field theories in Minkowski spacetime.

In the following, we demonstrate that the spacetime-dependent coefficients $m_5(x)$, $a_\mu(x)$, and $b_\mu(x)$ produce nontrivial corrections to the photon local dispersion relation. These modifications cause the vacuum to behave as an optically active medium, supporting nonpropagating modes typically associated with dichroism,  inducing the possibility of absorption and attenuation effects in electromagnetic wave propagation. To investigate these phenomena, we start from the $\nu=0$ component of Eq.\,\eqref{CovariantMEq2}, which allows us to write Gauss's law for the electric field as
\begin{align}
    \bm \na \cdot \bm E  =
    &\; \big(\bm \na c_1 \big) \cdot \bm E  
    + \big[\bm \na \big(c_2+ c_3\big)\big] \cdot  \bm B \,.
\end{align}
Similarly, 
from $\nu=i$ component of Eq.\,\eqref{CovariantMEq2} the Ampère-Maxwell equation in vacuum becomes
\begin{align} 
      \bm \na \times\bm B   - \pa_t \bm E   =&\; 
       \big(\bm \na c_1 \big) \times \bm B    - \big( \pa_t c_1\big)  \bm E - 
       \big[\bm \na \big(c_2+ c_3\big)\big]\times \bm E    
       - \big[ \pa_t \big(c_2+ c_3\big)\big]  \bm B \,.
\end{align}
As a consequence of the Bianchi identity, Gauss’s law for magnetism 
and Faraday-Lenz law remain unchanged:
\begin{align}
    &\bm \na \cdot \bm B = 0 , \\
    & \bm \na \times \bm E  + \pa_t \bm B = 0 \,. 
\end{align}

By taking the curl of Faraday-Lenz law and using the 
Ampère-Maxwell equation, the wave equation for the electric 
field takes the form
\begin{align} \label{WaveEq1}
    &\Big\{\Big[ \pa_t^2 - \bm \na^2 - \big( \bm \na c_1\big) \cdot \bm \na 
    - (\pa_t c_1) \pa_t  \Big] \de_{ij}  + \na_i \na_j 
    +  \big(\na_j \,c_1 \big) \na_i 
    \nn 
    \\
    &+ \varepsilon_{ijl} \Big[\na_l \big(c_2+ c_3\big)\pa_t 
    - \pa_t \big(c_2+ c_3\big) \na_l  \Big] \Big\} E_j = 0 .
\end{align}

There is a central complication in the modified wave equation 
\eqref{WaveEq1} related to the fact that the plane-wave ansatz 
(or the usual Fourier transform) is no longer the appropriate 
procedure to obtain the dispersion relation of the model, due 
to the presence of a background that depends explicitly on 
spacetime coordinates.
To proceed, we adopt a method commonly used in the plasma 
physics literature for non-homogeneous media, namely the 
eikonal approximation for electromagnetic waves \cite{Tracy14}, 
which allows one to define a local dispersion relation.

The eikonal approximation consists in the introduction of a small 
parameter $\epsilon^n$ multiplying each derivative term in the 
wave equation, where $n$ denotes the order of the derivative 
operator. This allows us to systematically track the different 
orders in a perturbative expansion that can be carried out 
afterward. At the end of the calculation, the limit $\epsilon \to 1$ 
is taken. Under these considerations, the eikonal approximation 
for the electric field is written as
\begin{align}
    E_i (t, \bm x) 
    = 
    A_i (t, \bm x) \exp \bigg[\dfrac{i}{\ep } S(t, \bm x)\bigg],
\end{align}
where the amplitude $A_i (t, \bm x)$ and the phase 
$S (t, \bm x)$ are both assumed to be real (see, e.g., 
Ref. \cite{Tracy14} for more details). Thus, at zeroth order in 
$\epsilon$, the wave equation is given by
\begin{align} \label{WaveEq2}
  M_{ij} (t, \bm x) E_j =0 ,
\end{align}
with
\begin{align}
  M_{ij} (t, \bm x)  =
  &\; \Big[ \omega^2 - \bm k^2 - i(\pa_t c_1) \omega 
  + i(\bm \na c_1)\cdot \bm k \Big] \de_{ij}  
  + k_i k_j - i ( \na_j \,c_1)  k_i 
  \nn 
  \\
  &\; +i \varepsilon_{ijl} \Big[\na_l \big(c_2+ c_3\big) \omega
  + \pa_t \big(c_2+ c_3\big) k_l \Big] .
\end{align}
and taking into account the definitions 
\begin{align}
  \na_i S(t,\bm x)  \equiv   k_i (t, \bm x) 
  \;\;\;\; \textrm{and }  \;\;\;\; 
  \pa_tS(t,\bm x) \equiv - \omega (t, \bm x)
\end{align}
as the local wave vector and the local frequency, respectively. 

Therefore, for a nontrivial electric field, the local dispersion 
equation (or eikonal equation) is obtained by 
$\mathcal{D} \big ( \omega 
= - \pa_t S,  \bm k
= \bm \nabla S, t, \bm x  \big ) \equiv  \det   M_{ij} (t, \bm x)=0$. 
Performing this determinant calculation and considering 
contributions up to second-order in LSV, one finds
\begin{align} \label{MDR}
   \mathcal{D} \big( \omega , \bm k, t,  \bm x  \big )  =
   &\;  
   4\om^6 
   -12i (\pa_t c_1)\om^5 
   - \Big[9\bm k^2 - 9i \bm k \cdot \big( \bm \na c_1\big) 
   +2 \big( \bm \na c_3\big)^2 \Big] \om^4 
   +2 \Big[9i\bm k^2 \big(\pa_t c_1 \big) 
    \nn
   \\
   &\;
   -2 \big( \pa_t c_3 \big) \bm k \cdot\big(\bm \na c_3 \big)  \Big] \om^3 
   + \Big\{ 2\bm k^2\Big[ 3\bm k^2 - \big(\pa_t c_3 \big)^2 
   + \big( \bm \na c_3\big)^2 -12 i \big( \bm \na c_1 \big) \cdot \bm k  \Big] 
    \nn
   \\
   &\;
   - \big[ \big( \bm \na c_3\big) \cdot \bm k\big]^2 \Big\} \om^2   
   -2 \bm k^2\Big[ 3i \bm k^2 \big(\pa_t c_1 \big) 
   - \big( \pa_t c_3\big) \big(\bm \na c_3 \big) \cdot \bm k \Big] \om 
   - \bm k^4 \Big[ \bm k^2 
    \nn
   \\
   &\;
   - \big( \pa_t c_3\big)^2  -3i \big(\bm \na c_1 \big) \cdot \bm k\Big] .
\end{align}

The modified dispersion relation \eqref{MDR} shows that the 
spacetime-dependent LSV contributions lead to a complex local 
frequency. The real part governs wave propagation, while the 
imaginary part, proportional to spacetime derivatives of $c_1$, may induce 
local amplification or attenuation of the wave amplitude. Indeed, by considering non-constant parameters means that the  energy-momentum of the wave is not conserved, by virtue of the exchange with the background, 
as discussed above. 
The wave may, therefore, acquire or lose energy.  When restricting the LSV sector to light-like structures, the dispersion relation becomes real, and only the parameter $m_5$ yields nontrivial contributions.

To close this Section, we wish to call the reader's attention to the fact that a spacetime-dependent frequency, as a consequence of non-constant/non-homogeneous LSV parameters, may lead to a red or a blue shift under specific conditions on the way the local parameters depend on time and space coordinates. We believe this a relevant issue to be inspected. The reason being that the LSV parameters are expected to originate from some more fundamental physics, valid at (high) energies close to the scale where the vacuum fluctuations are very strong. It is then reasonable to expect that the LSV entities should emerge as local parameters. Considering constant parameters is a very particular situation and, even though it is a particular case, a broad class of interesting effects beyond Standard-Model Physics may emerge.

\section{Concluding comments and prospects}
\label{sec5}

In this work, we present an explicit calculation of the effective photonic action up to fourth order in the electromagnetic field strength, taking  into account Lorentz symmetry violation effects up to second-order in the SME spacetime dependent  parameters $m_5$, $a_\mu$, and $b_\mu$. The calculation is carried out within the  proper-time formalism, with the heat-kernel results obtained from the evaluation  of the fermionic determinant via zeta-function regularization, following Ref.~\cite{McGar97}. 

At first-order in the electromagnetic field strength, an ultraviolet divergent contribution is generated, yielding a shift in the vacuum by virtue of the tadpole-type contribution to the effective action. At this  order, we also obtain a nontrivial finite contribution that may act as an external source to the Maxwell equations. In the case of a constant Lorentz symmetry-violating  background, all these terms reduce to surface terms and can therefore be dropped.

At second-order in the electromagnetic field strength, the corrections to the  effective action are entirely finite. The quadratic order is the only one that  yields a contribution at first-order in LSV arising from the $m_5$ term. This  correction leads to modified Maxwell equations in vacuum, which in turn result  in a non-conserved energy-momentum tensor for the photon field and a local  dispersion relation obtained within the eikonal approximation, which may encode potentially interesting phenomenology to be explored in a subsequent~work. 

The third order correction to the effective action in the field strength, when combined with the first-order contribution, points to a violation of Furry's  theorem by C-odd combinations at second-order in LSV. A similar result was previously  reported in Ref. \cite{Lane02} based on a diagrammatic analysis at first-order in LSV.

We have also worked out the contribution to the photonic action quartic in the electromagnetic field strengths and at the second-order in LSV parameters, as presented in Eq.\,\eqref{ac4ligh-light}. The resulting structures can be relevant in connection with the anomalous gauge boson interactions in the electroweak theory. In this regard, it is worth emphasizing that the neutral gauge boson, $Z^0$, also couples to fermionic currents via the weak charge, which yield, analogously, the emergence of neutral quartic vertices involving the $Z^0$ field~\cite{Kubota14,Kupco18}. From a phenomenological perspective, such interactions open up the possibility of constraining the associated parameters through photon-photon scattering measurements, particularly those reported by the ATLAS and CMS collaborations~\cite{ATLAS17,ATLAS19,CMS19}. Furthermore, we stress that the quartic operators appearing in Eq.\,\eqref{ac4ligh-light} correspond to contributions to light-by-light scattering already at the tree level.

Additionally, the attainment of an effective photonic action of the Euler-Heisenberg type including the space-time-dependent parameters may be of interest if we identify the four-vector, $a_\mu (x)$, and the pseudo-vector, $b_\mu (x)$, respectively with the gradients of the saxion (the CP-even axion field -- the supersymmetric scalar partner of the axion \cite{Kitano12, Roldán-Domínguez26}) and the axion field itself, in the context of Eq.\,\eqref{mazwellmod1}. This does not mean that these identifications reproduce interactions dictated by supersymmetry. We are merely borrowing the saxion field from the supermultiplet that accommodates the axion in a supersymmetric scenario. This could be a way to introduce dark matter candidates in interaction with the photon in a sort of extended Euler-Heisenberg action. By upgrading these variable LSV parameters to the status of dynamical fields, one should also introduce their respective kinetic and mass terms. If we take this path, an effective axionic Euler-Heisenberg emerges which may be further exploited. Differently from simply adding the (purely photonic) Euler-Heisenberg action to the Axionic Electrodynamics, such a construction would incorporate the saxion and axion fields into the classes of terms with powers of the photon field-strength and its dual.

To conclude, we emphasize, upon comparison with the results of Ref. \cite{Ferrari21}, that there is an  explicit dependence on the parametrization adopted to construct the quadratic  Dirac operator in Eq.\,\eqref{quadraticO}, which affects the final form of the effective action.  We briefly argue that this ambiguity may be understood in terms of the  multiplicative anomaly, which reflects the breakdown of the naive factorization  property of functional determinants under a regularization procedure. This issue will be addressed in detail in a forthcoming work, where we consider  constant $b_\mu$ and $H_{\mu\nu}$ terms. In that analysis, we shall clarify the  origin of this ambiguity and argue in favor of the parametrization involving the  $\ga_5$ matrix as the most reliable one, showing that it is consistent with  results obtained through Feynman diagram calculations.

\section*{Acknowledgments}

 The authors are grateful to Mirian Reetz for useful comments. W.C.S. is grateful for the financial support from \textit{Conselho Nacional de Desenvolvimento Cient\'{i}fico e Tecnol\'{o}gico} (CNPq - Brazil) for the Postdoctoral Fellowship [PCI Grant No. 314125/2025-6]. J.P.S.M. gratefully acknowledges financial support from the Fundação Carlos Chagas Filho de Amparo à Pesquisa do Estado do Rio de  Janeiro (FAPERJ) under Grant No. E-03/203.638/2024 (Doutorado  Nota 10 Fellowship).
 
 \section*{Data Availability}
 
 \noindent
 There are no publicly available research data or software supporting 
 this manuscript.
 Requests for further information or data should be sent to the authors.
 

\appendix
\section*{Appendix}
\addcontentsline{toc}{section}{Appendices}
\renewcommand{\thesubsection}{\Alph{subsection}}
\subsection{Wick rotation conventions} 
\label{App}

The calculation of the effective action performed in Sec.~\ref{sec3} 
is well defined in Euclidean space, $\mathbb{E}^4$. In this Appendix, 
we summarize the conventions adopted and the corresponding 
relations used to implement the Wick rotation from Minkowski 
spacetime, $\mathbb{M}^{1,3}$, with metric tensor 
$\eta_{\mu\nu} = \textrm{diag}(1,-1,-1,-1)$, to Euclidean space 
$\mathbb{E}^4$, endowed with the metric 
$\hat g_{\mu\nu} = - \de_{\mu\nu}$.
In $\mathbb{M}^{1,3}$, Greek indices run over $\mu, \nu = (0,1,2,3)$, 
whereas in $\mathbb{E}^4$ they run over $\mu, \nu = (1,2,3,4)$.

At the outset, the conventions adopted for the Wick rotation affect 
the coordinates $x^\mu$, the partial derivatives $\partial_\mu$, and 
the four-vectors $v^\mu$ defined in $\mathbb{M}^{1,3}$. The 
correspondence between these quantities in $\mathbb{M}^{1,3}$ and 
their counterparts in $\mathbb{E}^4$ is given, respectively, by
\begin{align}
& \; \{x^0 \to -i \hat x_4, 
   \quad 
   x^i \to \hat{x}_i \},
\\
& \; \{\pa_0 \to i\hat{\pa}_4, 
    \;\quad 
    \na_i \to \hat{\na}_i \},
\\
& \; \{v^0 \to -i\hat{v}_4,
 \;\quad
 v^i \to \hat{v}_i \},
\end{align}
where the bar denotes quantities defined in Euclidean space. These 
definitions imply that the scalar product of four-vectors is transformed 
in the following manner
\begin{align}
    u_\mu v^\mu 
    &
    \to
    - \hat u_\mu \hat v^\mu ,
    \\
     \pa_\mu v^\mu 
     &
     \to
     \hat \pa_\mu \hat v^\mu.
\end{align}
For the Dirac gamma matrices, we adopt the following conventions:
\begin{align}
   \{ \ga^i \to  i   \hat {\ga}^i; \quad \ga^0 \to  \hat{\ga}^4\},
\end{align}
with the Euclidean gamma matrices taken to be Hermitian, i.e.,
$\hat \ga^{i\dagger}= \hat\ga^i$ 
and 
${\hat \ga}^{4\dagger}= \hat{ \ga}^4$.
The consequences of these conventions for the Clifford algebra are
\begin{align}
    \{ \ga^\mu , \ga^\nu\}  = 2\eta^{\mu\nu}
    \to  
    \{ \hat \ga^\mu ,\hat  \ga^\nu\} = 2\hat{g}^{\mu\nu}=- 2 \de^{\mu\nu} .
\end{align} 
In this case, the matrix $\gamma_5$ transforms accordingly as
\begin{align}
    \ga_5 = - i \ga^0 \ga^1 \ga^2\ga^3  
     \to    
     {\hat \ga}^1 {\hat \ga}^2 {\hat \ga}^3 {\hat \ga}^4 =   \hat \ga_5.
\end{align}
It is straightforward to verify that $\hat{\gamma}_5$ is  Hermitian, 
$\hat{\gamma}_5^\dagger = \ga_5 $. The contraction of gamma 
matrices with the derivative operator and an arbitrary four-vector 
follows the rules
\begin{align}
    &
    \ga^\mu \pa_\mu \to  i \hat \ga^\mu \hat \pa_\mu ,
    \\
   & 
   v_\mu \ga^\mu \to - i \hat{v}_\mu \hat{\ga}^\mu.
\end{align}
Finally, we show how the components of the Lorentz group generator
$\Sigma^{\mu\nu} = \tfrac{i}{4}[\gamma^\mu,\gamma^\nu]$ transform 
according to the prescriptions above. This yields
\begin{align}
    \Si^{0i} \to i   \hat \Si^{4i},
    \quad
    \Si^{ij} \to -  \hat \Si^{ij}.
\end{align}
Thus, the contraction of two arbitrary four-vectors $u_\mu$ and $v_\nu$ 
with $\Sigma^{\mu\nu}$ transforms as
\begin{align}
   u_\mu v_\nu \Sigma^{\mu\nu} \to -\hat{u}_\mu \hat{v}_\nu \hat \Si^{\mu\nu} .  
\end{align}
For the electromagnetic field-strength tensor, we have
\begin{align}
     F_{0i} \to -i \hat F_{4i}, 
     \quad 
     F^{ij} \to  -\hat F^{ij},
\end{align}
which implies that $F_{\mu\nu}^2= \hat F_{\mu\nu}^2$.

To determine the correspondence for terms involving the dual 
electromagnetic field-strength tensor, it is necessary to examine how the 
four-dimensional Levi-Civita symbol transforms. We adopt the conventions 
$\varepsilon^{0123} = +1$ and $\varepsilon_{0123} = -1$ for the Levi-Civita 
symbol in Minkowski spacetime. Under the Wick rotation, the corresponding 
relations in Euclidean space are
\begin{align}
    &
    \varepsilon^{0123} = +1 \to i \hat \varepsilon^{1234} = +i, 
    \\
    &
    \varepsilon_{0123} =-1 \to -i \hat \varepsilon_{1234} = -i.
\end{align}
With these definitions of the Levi-Civita symbol, we can establish the mapping 
of the dual field-strength tensor, 
$\tilde F^{\mu\nu} = \tfrac{1}{2} \varepsilon^{\mu\nu\ka\la} F_{\ka\la}$, from 
Minkowski spacetime $\mathbb{M}^{1,3}$ to Euclidean space $\mathbb{E}^{4}$. 
In particular, the components transform as
\begin{align}
   \tilde F_{0i} \to -\hat{\tilde{F}}_{4i}, 
   \quad
   \tilde F_{ij} \to - i \hat{\tilde{F}}_{ji}. 
   \label{dual3}
\end{align}
From these relations, it follows that 
$\tilde{F}_{\mu\nu} F^{\mu\nu } \to i \bm  \hat{\tilde{F}}_{\mu\nu} \hat F^{\mu \nu }$.
We can also determine the correspondence of contractions involving the Lorentz 
generators $\Sigma^{\mu\nu}$ and the tensors $F_{\mu\nu}$ and 
$\tilde{F}_{\mu\nu}$:
\begin{align}
    &
    \Sigma^{\mu\nu} F_{\mu\nu} 
    \to 
    \hat \Sigma^{\mu\nu} \hat  F_{\mu\nu} , 
    \\
   & 
  \Sigma^{\mu\nu} \tilde F_{\mu\nu} 
  \to 
  -i \hat \Sigma^{\mu\nu} \hat{\tilde{F}}_{\mu\nu} .
\end{align}

Finally, according to these conventions, the components of the covariant derivative
$D_\mu = \partial_\mu + i e A_\mu$ transforms as
\begin{align}
\{D_0 \to i \hat D_4, 
\quad 
D_i \to \hat {D}_i \}.
\end{align}
Thus, contractions involving covariant derivatives take the form
\begin{align}
    &
    \ga^\mu D_\mu 
    \to 
    i \hat \ga^\mu \hat D_\mu,
\\
    &
    D_\mu v^\mu 
    \to
     \hat D_\mu \hat v^\mu . 
\end{align}

These are the main conversions 
adopted in the present work. Obviously, the same relations apply in the inverse 
Wick rotation to express the final results of Sec.~\ref{sec3} in Minkowski spacetime.

\end{document}